\documentclass[11pt]{article}
\usepackage{moriond,epsfig}

\def\mco{\multicolumn}

\begin{document}
\vspace*{4cm}
\title{POLARIZATION STUDY IN B DECAYS TO VECTOR FINAL STATES}

\author{ CAI-DIAN L\"U }

\address{Institute of High
Energy Physics, P.O.Box 918(4), Beijing 100049, China}

\maketitle \abstracts{ The small longitudinal polarization fractions
(50\%) of $B \to \phi K^*$ measured by B factories contradict with
the naive theoretical counting rules. We review the current
theoretical status of the $B\to VV$ decay studies and calculate many
of them in the perturbative QCD factorization approach based on
$k_T$ factorization. We find that the penguin annihilation and
non-factorizable emission diagrams can enhance the transverse
polarization fractions. The PQCD results agree with experiments for
the measured $B\to \phi K^*$, $B\to \rho K^*$ and $B\to \rho \rho$
channels, and we also predict new results (some different from other
approaches) for those not yet measured channels.}

\section{Introduction}

The abnormally large transverse momentum fraction measured by the B
factories in the $B\to K^* \phi$ decays \cite{phike} arouse many
discussions in the framework of standard model with hadronic
uncertainties and also new physics contributions
\cite{kagan,phik,scet,fsi}. Among these explanations some still face
problem for explaining the identity fraction of the two transverse
polarizations or the large relative strong phase between
polarizations. In fact the perturbative QCD factorization approach
(PQCD) based on $k_t$ factorization can really do a good job with
59\% of the longitudinal polarization fraction and also the right
ratio of the two transverse polarizations and right strong
phases\cite{kphi-pqcd}. The reason is that in the PQCD approach, a
not very small space like penguin annihilation diagram contribute
largely for the transverse polarizations. This annihilation type
diagrams also contribute a large strong phase\cite{pipi,pik}. Not
surprisingly, the recent direct CP measurements of two B factories
in $B^0 \to \pi^+ \pi^-$, $\pi^- K^+$ decays also agree with the
previous PQCD predictions \cite{direct}

Inspired by the successful achievement for the PQCD framework, we
studied most of the charmless decays of B meson with two vector
final states and also some $B_s$ decay channels. We find that
longitudinal polarization fraction of those penguin dominant decays
are indeed suppressed by the space like penguin diagrams due to
(S-P)(S+P) operators.

\section{perturbative QCD approach formalism}

In non-leptonic B decays, it is the heavy b quark decay through
electroweak interaction, usually interchanging a W boson. By loop
diagrams penguin operators are also involved, which together make
the effective Hamiltonian for the weak decays:
\begin{equation}
\label{heff} {\cal H}_{{\it eff}} = \frac{G_{F}} {\sqrt{2}} \,
\left[ V_{ub} V_{us}^* \left (C_1 O_1^u + C_2 O_2^u \right) -
V_{tb} V_{ts}^* \, \left(\sum_{i=3}^{10} C_{i} \, O_i + C_g O_g
\right) \right] \quad ,
\end{equation}
where
\begin{equation}\begin{array}{llllll}
 O_1^{u} & = & \bar s_\alpha\gamma^\mu L u_\beta\cdot \bar
u_\beta\gamma_\mu L b_\alpha\ , &O_2^{u} & = &\bar
s_\alpha\gamma^\mu L u_\alpha\cdot \bar
u_\beta\gamma_\mu L b_\beta\ , \\
O_3 & = & \bar s_\alpha\gamma^\mu L b_\alpha\cdot \sum_{q'}\bar
 q_\beta'\gamma_\mu L q_\beta'\ ,  &
O_4 & = & \bar s_\alpha\gamma^\mu L b_\beta\cdot \sum_{q'}\bar
q_\beta'\gamma_\mu L q_\alpha'\ , \\
O_5 & = & \bar s_\alpha\gamma^\mu L b_\alpha\cdot \sum_{q'}\bar
q_\beta'\gamma_\mu R q_\beta'\ ,  & O_6 & = & \bar
s_\alpha\gamma^\mu L b_\beta\cdot \sum_{q'}\bar
q_\beta'\gamma_\mu R q_\alpha'\ , \\
O_7 & = & \frac{3}{2}\bar s_\alpha\gamma^\mu L b_\alpha\cdot
\sum_{q'}e_{q'}\bar q_\beta'\gamma_\mu R q_\beta'\ ,  & O_8 & = &
\frac{3}{2}\bar s_\alpha\gamma^\mu L b_\beta\cdot
\sum_{q'}e_{q'}\bar q_\beta'\gamma_\mu R q_\alpha'\ , \\
O_9 & = & \frac{3}{2}\bar s_\alpha\gamma^\mu L b_\alpha\cdot
\sum_{q'}e_{q'}\bar q_\beta'\gamma_\mu L q_\beta'\ ,  & O_{10} & =
& \frac{3}{2}\bar s_\alpha\gamma^\mu L b_\beta\cdot
\sum_{q'}e_{q'}\bar q_\beta'\gamma_\mu L q_\alpha'~.
\label{operators}
\end{array}
\end{equation}
Here $\alpha$ and $\beta$ are the $SU(3)$ color indices; $L$ and $R$
are the left- and right-handed projection operators with $L=(1 -
\gamma_5)$, $R= (1 + \gamma_5)$. The sum over $q'$ runs over the
quark fields that are active at the scale $\mu=O(m_b)$, i.e.,
$(q'\epsilon\{u,d,s,c,b\})$. For $b\to d$ transitions, one need only
replace the $s$ quark with d quark in eq.(\ref{operators}).

The effective four quark operators describe the hard electroweak
process in b quark decays, however hadronization is needed for the
meson decays. In hadronic B decays, more than one energy scale is
involved, the factorization technique is very important here. Since
all the decays are electro-weak decays, the electro-weak breaking
scale 100GeV is involved. Unavoidably, the hadronization scale
200MeV is for the hadronic decays, which is non-perturbative. In the
intermediate scale, the b quark mass is the energy release scale in
these decays. Therefore a factorization theorem is required for the
at least three energy scales.

For B meson decays with two light vector mesons in the final states,
the light mesons obtain large momentum of 2.6GeV in the B meson rest
frame. All the quarks inside the light mesons are therefore
collinear like. Since the heavy b quark in B meson carry most of the
energy of B meson, the light quark in B meson is soft. In the usual
emission diagram of B decays, this quark goes to the final state
meson  without electroweak interaction with other quarks, which is
called a spectator quark. Therefore there must be a connecting hard
gluon to make it from soft like to collinear like.
   The hard part of the interaction
becomes six quark operator rather than four. The soft dynamics here
is factorized into the meson wave functions. The decay amplitude is
infrared safe and can be factorized as the following formalism:
\begin{equation}
C(t) \times H(t) \times \Phi (x) \times \exp\left[ -s(P,b) -2 \int
_{1/b}^t \frac{ d \bar\mu}{\bar \mu} \gamma_q (\alpha_s (\bar
\mu)) \right], \label{eq:factorization_formula}
\end{equation}
where $C(t)$ are the corresponding Wilson coefficients of four quark
operators, $\Phi (x)$ are the meson wave functions and the variable
$t$ denotes the largest energy scale of hard process $H$, which is
the typical energy scale in PQCD approach and the Wilson
coefficients are evolved to this scale.  The exponential of $S$
function is the so-called Sudakov form factor resulting from the
resummation of double logarithms occurred in the QCD loop
corrections, which can suppress the contribution from the
non-perturbative region, making the perturbative region to give the
dominant contribution. The ``$\times$'' here denotes convolution,
i.e., the integral on the momentum fractions and the transverse
intervals of the corresponding mesons. Since logarithm corrections
have been summed by renormalization group equations, the
 above factorization formula does not depend on the renormalization
scale $\mu$ explicitly.

\section{$B\to VV$ decays in the PQCD approach}

\begin{figure}
\psfig{figure=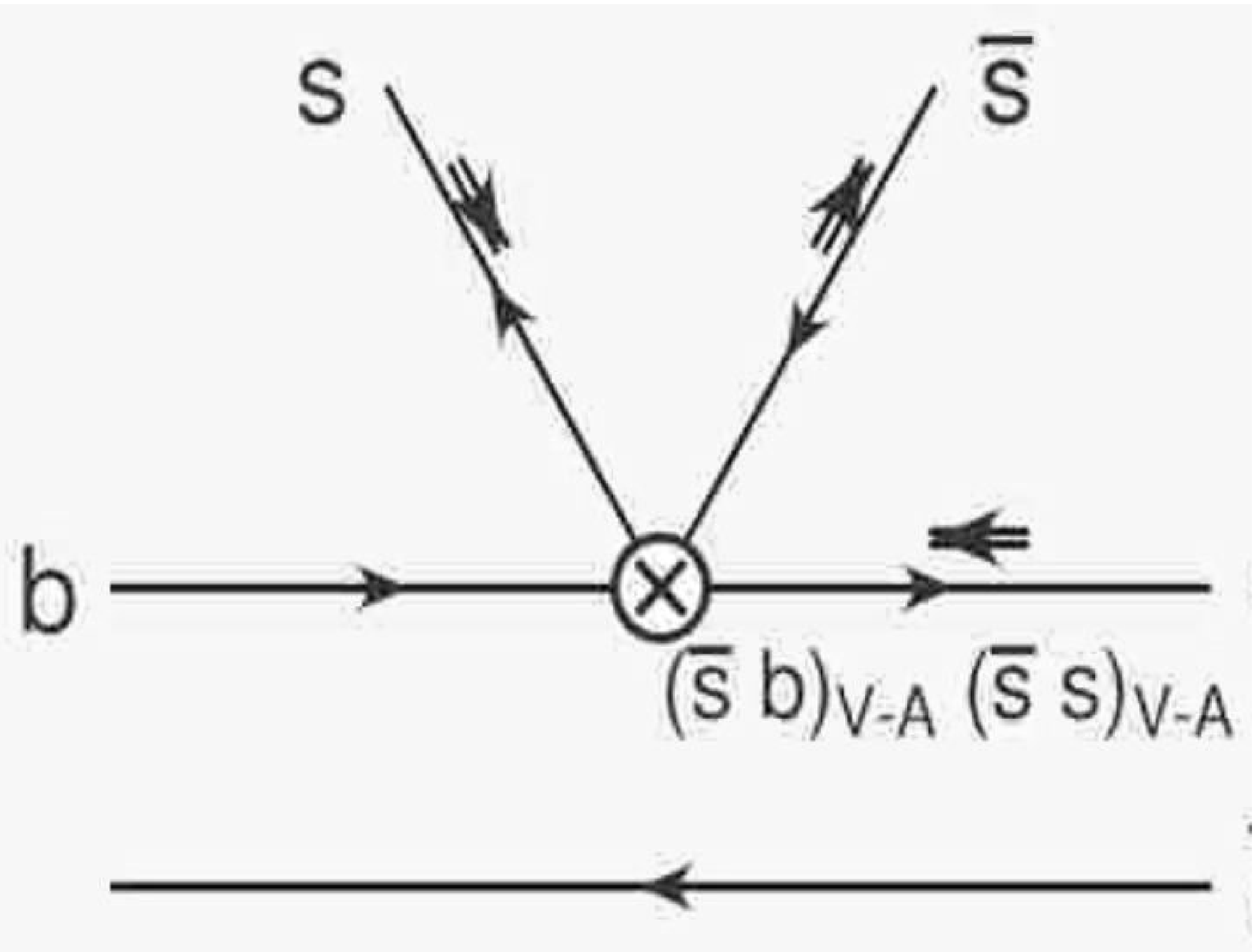,height=1.in}\hfill
\psfig{figure=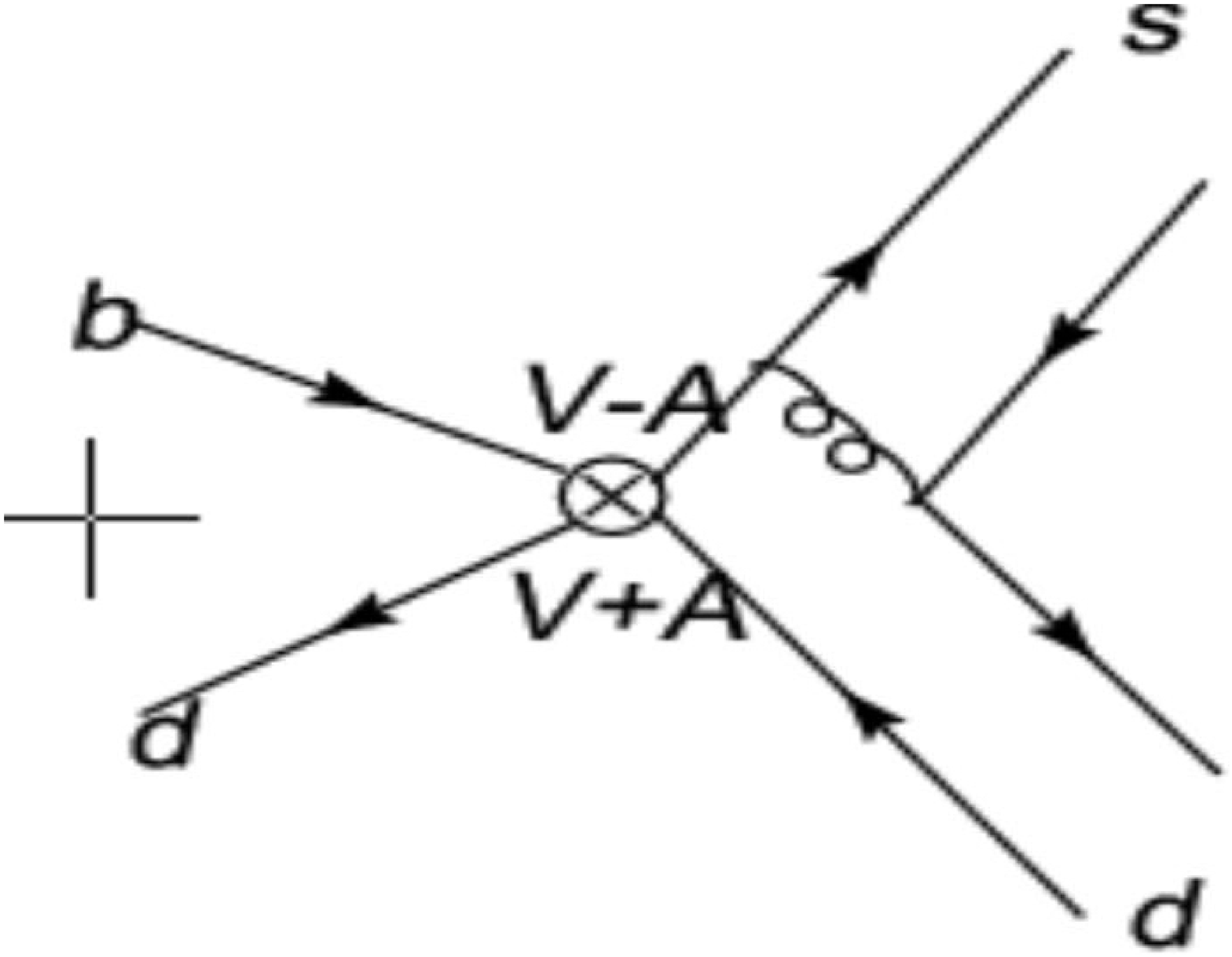,height=1.in}\hfill
\psfig{figure=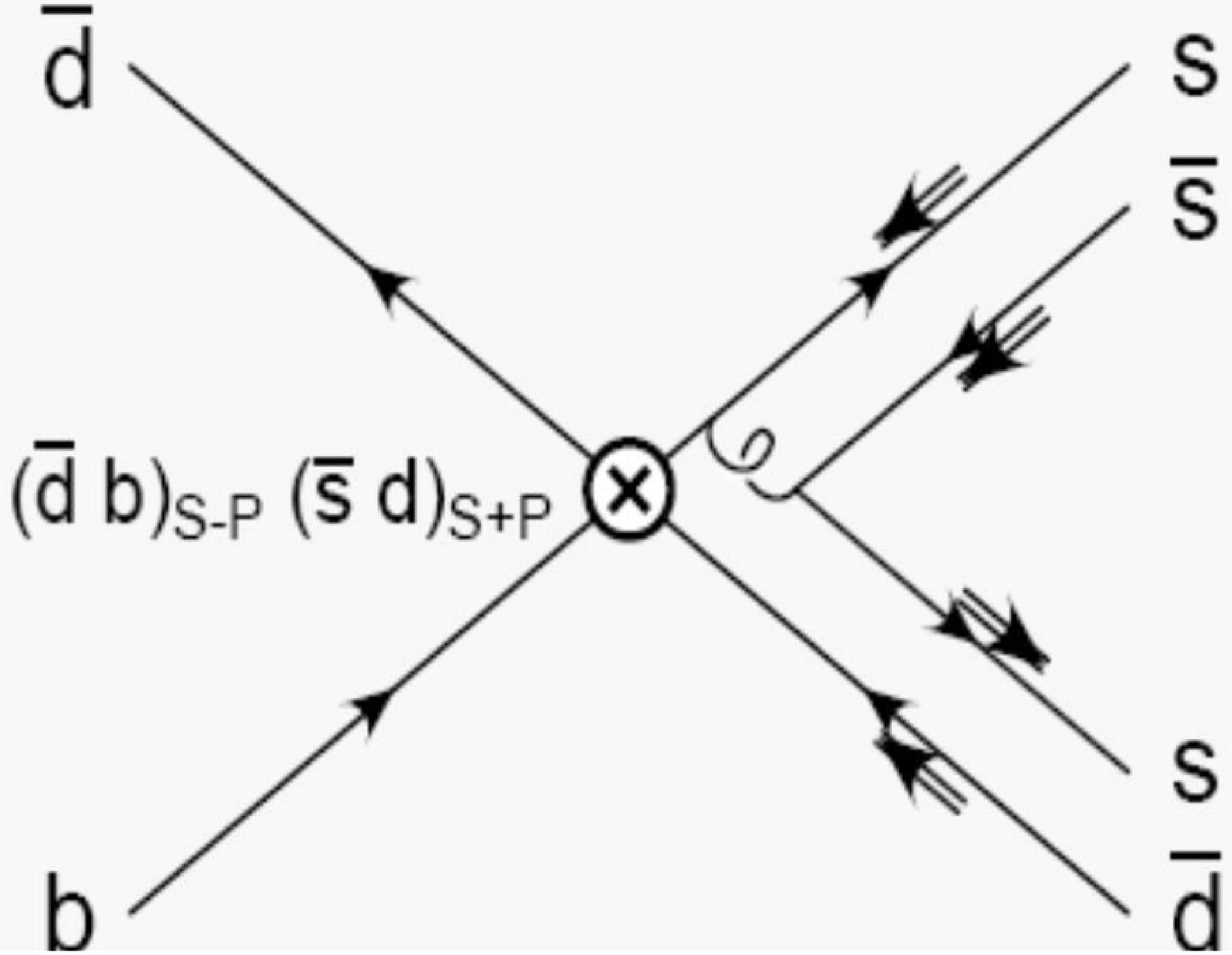,height=1.in}
\\
\centerline{\small (a) \hspace{5cm}\small (b) \hspace{5cm}\small
(c)} \caption{Feynman diagrams contributing to $B\to VV$ decays:
spectator diagram (a); annihilation diagram (b);  annihilation
diagram after Fiertz transformation (c). \label{fig}}
\end{figure}

In standard model, the four quark operators in eq.(\ref{operators})
is either (V-A) or (V+A), which implies that the emitted meson with
left-handed quark and right-handed unti-quark or the inverse case.
This spin structure is shown in Fig.\ref{fig}(a). Therefore to make
a longitudinal polarization meson costless, but require at least one
quark spin flip to make a transverse polarized meson. Since the
quark spin flip is suppressed in the heavy quark limit, the
transverse polarization is suppressed in the charmless B meson
decays. But for the space like penguin annihilation diagrams with
operator $O_6$, the situation shown in Fig.\ref{fig}(b) is different
\cite{kagan}. This operator is $(V-A)(V+A)$ structure which becomes
(S+P)(S-P) after Fiertz transformation, shown in Fig.\ref{fig}(c).
In this case, the produced quark unti-quark pair going to different
mesons contribute to the three polarizations almost equally, thus
the transverse polarization gets twice contribution of longitudinal
one. The annihilation contribution of operator $O_6$ is  chirally
enhanced in PQCD approach, therefore we can have a larger transverse
contribution in penguin dominant decays like $B\to K^*\phi$
\cite{kphi-pqcd}, while in QCD factorization approach, one has to
increase the annihilation contribution by hand \cite{kagan}.

The input wave functions and various parameters are shown in the
corresponding papers \cite{kprho,rhorho,phiphi,kpkp}. The numerical
results for some of the $B\to K^*\rho$, $\rho(\omega) \rho(\omega)$
decays are shown in table~\ref{tab1} together with some of the
experimental measurements\cite{phike,rhorho-e}.
 From the table, one can see that the branching ratios calculated
 by PQCD approach agree well with the experiments. As for the
 measured polarization fractions most of them agree well except
 for $B^+ \to\rho^+K^{*0}$, where there is a large discrepancy between
 the two experiments \cite{phike}, our results agree with BABAR.
The uncertainty showed in the table for $B\to \rho K^*$ decays are
only from the change of $K^*$ wave functions, which shows high
sensitivity of results with meson wave functions \cite{kprho}. The
tree dominant $B\to \rho^+ \rho^-$, $\rho^+\rho^0$ and
$\rho^+\omega$ decays are indeed longitudinal polarization dominant
(with more than 90\%). Meanwhile, the penguin dominant decays $B\to
K^*\rho (\omega)$ have a reasonable transverse polarization fraction
mainly due to a non-negligible annihilation diagram contribution.

\begin{table}[th]
\begin{center}\caption{ Branching ratios ($10^{-6}$)
and  polarization fractions using different type of light meson
wave functions (the CKM phase angle $\phi_3$ is fixed as
$60^{\circ}$)}
\begin{tabular}{|c|c|c|c|c|c|c|}
\hline\hline
     & \mco{2}{|c|}{ Branching ratio} & \mco{2}{|c|}{polarization fraction $R_L(\%)$} &&\\
     \cline{2-5}
 Decay  & theory &   exp. & theory  &exp. & $R_{\parallel }(\%)$ & $R_{\perp} (\%)$ \\
   \hline\hline
$B^0\to\rho^- K^{*+ }$&  10-13  & $ \le $ 24   & $71-78$ & &12 &10
 \\ \hline
 $ B^+\to\rho^+ K^{*0 }$& 13-17 & $10.5\pm 1.8$   &$76-82$& $66
 \pm 7$ &13&10
\\ \hline

$B^+\to\rho^0 K^{*+ }$ &6-9 &$10.6^{+3.8}_{-3.5}$  &$78-85$ &
$96^{+4}_{-15}\pm 4 $ &11&11
\\ \hline
 $  B^+\to\omega K^{*+ }$&5-8 &$<7.4$& $73-81$& &19&9
\\
\hline $B^0 \to \rho^+ \rho^-   $  & $35 \pm 5\pm 4$          &
$30\pm 6$
 & 94 &$96^{+4}_{-7}$& 3     & 3            \\\hline
$B^+ \to \rho^+ \rho^0   $  & $17 \pm 2\pm 1$          &
$26.4^{+6.1}_{-6.4}$
 & 94 &$99\pm5$  & 4     & 2            \\\hline
$B^+ \to \rho^+ \omega   $  & $19 \pm2\pm1$           &
$12.6^{+4.1}_{-3.8}$
 & 97 &$88^{+12}_{-15}$  & 1.5   & 1.5          \\\hline
$B^0 \to \rho^0 \rho^0   $  & $0.9  \pm0.1\pm0.1$         & $<1.1$
   & 60 & - & 22    & 18           \\\hline
$B^0 \to \rho^0 \omega   $  & $1.9 \pm0.2\pm0.2$         & $<3.3$
               & 87 & - & 6.5   & 6.5          \\\hline
$B^0 \to \omega \omega   $  & $1.2\pm0.2\pm0.2 $         &  $<19$
            & 82 & - & 9     & 9            \\
            \hline\hline
\end{tabular}\label{tab1}
\end{center}
\end{table}

There are also time-like penguin contribution dominant decay
channels such as $B^0 \to \phi \phi$\cite{phiphi}, $B\to K^* K^*$
decays\cite{kpkp} etc. The perturbative QCD factorization approach
calculation shows that reasonable transverse polarization fractions
are about 30\% in these decays. However, being CKM parameter
$|V_{tb}V_{td}^*|$ suppressed, their branching ratios are $10^{-8} -
10^{-7}$, which are very difficult to be measured.

With the coming LHCb experiments, a large number of $B_s$ and $B_c$
mesons can be produced. We study the $B_s\to \rho K^*$
decays\cite{bskprho}, with $B_s$ meson wave function constrained
from other $B_s$ decays\cite{}. The tree dominant mode $B_s\to
\rho^+K^{*-}$ is indeed longitudinal polarization dominant with more
than 90\%, while the color suppressed modes with only 40\%
longitudinal polarization. This is similar with the $B\to \rho \rho
(\omega)$ case\cite{rhorho}, which will provide a further test of
the theory.

\section{Summary}

The polarization fractions measured by the two B factory experiments
provide a test for various theories in the non-leptonic B decays.
The perturbative QCD factorization approach based on $k_T$
factorization can explain the polarization fractions without new
input parameters. The space like penguin annihilation diagrams,
which is the main source of strong phase to explain the direct CP
measurement of B decays, play an essential role in the enhancement
of the transverse polarization fractions.

\section*{Acknowledgments}

The author would like to thank Y. Li, Y.L. Shen, W. Wang and J.
Zhu for contributions on works related here.
 This work was partly supported by
the National Science Foundation of China.

\end{document}